\documentclass[conference, twocolumn]{IEEEtran}

\usepackage[utf8]{inputenc}
\usepackage{microtype}

\usepackage{amssymb}
\usepackage[abs]{overpic}
\usepackage{graphicx}
\usepackage{subcaption}
\usepackage{siunitx}
\usepackage[hyphens]{url}
\usepackage{cite}

\usepackage{paralist}
\usepackage{enumerate}
\usepackage{enumitem}
\usepackage{algorithm}
\usepackage{xspace}
\usepackage[noend]{algpseudocode}

\usepackage{booktabs}
\usepackage[table,xcdraw,dvipsnames]{xcolor}

\usepackage[flushleft]{threeparttable}

\usepackage{wasysym}%

\usepackage{balance}
\usepackage{dcolumn}
\usepackage{longtable,tabu}
\usepackage{threeparttablex}
\usepackage{adjustbox}
\usepackage{tabularx}
\usepackage{tikz}
\usepackage{pgfplots}
\pgfplotsset{width=6.5cm,compat=1.9}
\usepgfplotslibrary{statistics}

\usepackage[underline=true]{pgf-umlsd}
\usetikzlibrary{calc}
\usetikzlibrary{arrows.meta}

\usepackage{arydshln}
\usepackage{multirow}

\usepackage{pifont}
\usepackage{enumitem}

\begin{document}
\title{E-Voting with Blockchain: An E-Voting Protocol with Decentralisation and Voter Privacy}
\author{\IEEEauthorblockN{Freya Sheer Hardwick, Apostolos Gioulis, Raja Naeem Akram, and Konstantinos Markantonakis}
\IEEEauthorblockA{ISG-SCC, Royal Holloway, University of London, Egham, United Kingdom\\
Email: \{Freya.SheerHardwick.2016, Apostolos.Gioulis.2015\}@live.rhul.ac.uk, \{r.n.akram, k.markantonakis\}@rhul.ac.uk}}

\maketitle

\begin{abstract}
Technology has positive impacts on many aspects of our social life. Designing a 24 hour globally connected architecture enables ease of access to a variety of resources and services. Furthermore, technology like the Internet has been a fertile ground for innovation and creativity. One such disruptive innovation is blockchain -- a keystone of cryptocurrencies. The blockchain technology is presented as a game changer for many of the existing and emerging technologies/services. With its immutability property and decentralised architecture, it is taking centre stage in many services as an equalisation factor to the current parity between consumers and large corporations/governments. One potential application of the blockchain is in e-voting schemes. The objective of such a scheme would be to provide a decentralised architecture to run and support a voting scheme that is open, fair, and independently verifiable. In this paper, we propose a potential new e-voting protocol that utilises the blockchain as a transparent ballot box. The protocol has been designed to adhere to fundamental e-voting properties as well as offer a degree of decentralisation and allow for the voter to change/update their vote (within the permissible voting period). This paper highlights the pros and cons of using blockchain for such a proposal from a practical point view in both development/deployment and usage contexts. Concluding the paper is a potential roadmap for blockchain technology to be able to support complex applications.
\end{abstract}

\IEEEpeerreviewmaketitle

\section{Introduction}
\label{sec:Introduction}
Voting, whether traditional ballet based or electronic voting (e-voting), is what modern democracies are built upon. In recent years voter apathy has been increasing, especially among the younger computer/tech savvy generation \cite{VoterApathy2005}. E-voting is pushed as a potential solution to attract young voters \cite{eggers2007government,harrison2012creating}. For a robust e-voting scheme, a number of functional and security requirements are specified \cite{wang2017review,gritzalis2002principles,anane2007voting} including transparency, accuracy, auditability, system and data integrity, secrecy/privacy, availability, and distribution of authority. 

Blockchain technology is supported by a distributed network consisting of a large number of interconnected nodes. Each of these nodes have their own copy of the distributed ledger that contains the full history of all transactions the network has processed. There is no single authority that controls the network. If the majority of the nodes agree, they accept a transaction. This network allows users to remain anonymous. A basic analysis of the blockchain technology (including smart contracts) suggests that it is a suitable basis for e-voting and, moreover, it could have the potential to make e-voting more acceptable and reliable. There are number of papers that have explored this idea \cite{Moura:2017,ayed2017conceptual,mccorry2017smart} including now this one.

Obvious advantages of e-voting using blockchains includes: i) greater transparency due to open and distributed ledgers, ii) inherent anonymity , iii) security and reliability (especially against Denial of Service Attacks) and iv) immutability (strong integrity for the voting scheme and individual votes). Existing works explore how blockchains can be used to improve the e-voting schemes or provide some strong guarantees of the above listed requirements. However, these papers do not discuss the implementation challenges and limitations of the blockchain (and smart contract) technologies at their current state to fully support a large scale voting scheme. In this paper we explore both the possibilities of an e-voting scheme, along with the challenges and limitation of the blockchain technology in the e-voting context. 

\subsection{Contribution of the Paper}
\label{sec:COntributionofthePaper}
Contributions of the paper can be summed up as below:

\begin{enumerate}
\item The paper proposes an e-voting scheme based on blockchain technology that meets the fundamental e-voting properties whilst, at the same time, provides a degree of decentralisation and places as much control of the process in the hands of the voters as was deemed possible.
\item Discussion on the implementation challenges and underlying platform's (blockchain and smart contracts) limitation to support the e-voting proposal.
\end{enumerate}

A short discussion about e-voting schemes is provided in section \ref{sec:eVoting}. Protocol used as part of the proposed e-voting scheme is analysed in section \ref{sec:ProposedProtocol}. Implementation and operation evaluation is provided in section \ref{sec:ImplementationandOperationEvaluation}.

\section{E-Voting}
\label{sec:eVoting}
Voting mechanisms using electronic means, or 'e-voting mechanisms', to aid casting and counting votes have been studied in both the commercial and the academic world. 


\subsection{Design Properties}
In order for an e-voting protocol to be deemed secure certain formally-stated properties must hold. 

\begin{itemize}
\item Fairness: No early results should be obtainable before the end of the voting process; this provides the assurance that the remaining voters will not be influenced in their vote.

\item Eligibility: This property states that only eligible voters should be allowed to cast their vote and they should do so only once. The basis of this property is authentication, since voters need to prove their identity before being deemed eligible or not. 

\item Privacy: The way that an individual voter voted should not be revealed to anyone. This property in non-electronic voting schemes is ensured by physically protecting the voter from prying eyes. 

\item Verifiability: This property guaranties that all parties involved have the ability to check whether their votes have been counted or not. Typically two forms of verifiability are defined, individual and universal verifiability. Individual verifiability gives an individual voter the ability to verify that one’s vote has been counted. Universal verifiability requires that anyone can verify that the election outcome is the one published. 

\item Coercion-resistance: A coarser should not have the ability to distinguish whether a coerced voter voted the way they were instructed to.
\end{itemize}

It is within the scope of the paper to conceive of a protocol that has the aforementioned properties. However Coercion resistance will not be actively pursued since it was deemed not possible to be achieved purely with technological means in a remote e-voting protocol. The protocol does however have the property of Forgiveness that can be perceived as a weaker notion of the coercion-resistance property. 

\begin{itemize}
\item Forgiveness: The ability of a voter to alter ones vote after it has been cast. This property links to coercion resistance because it provides a coerced voter the option of changing ones vote at a later stage in order to reflect ones true opinion. 
\end{itemize}

\subsection{Blockchain e-voting}
\label{sec:BlockchainEVoting}
The utilisation of Blockchain technology in e-voting applications is not a new thing. Many schemes have been proposed; however, most of those protocols lack proper documentation and questions remain about their internal workings. Table I, gives an overview of the degree to which three of the most well documented and popular, commercial, remote e-voting protocols, namely BitCongress \cite{BItCongress}, Follow my Vote \cite{FollowMyVote} and TIVI \cite{TIVI}, satisfy the fundamental e-voting properties.

\begin{table}
\begin{minipage}[b]{0.5\textwidth}
\centering
\begin{tabularx}{\textwidth}{|X | X | X | X|}
\hline
\multirow{2}{*}{\emph{Properties}}& \multicolumn{3}{|c|}{\emph{Protocols}} 
\tabularnewline
\cline{2-4} 
& \emph{Bitcongress} & \emph{Follow My Vote} & \emph{TIVI} 
\tabularnewline
\hline
Fairness & No & No & No 
\tabularnewline
\hline
Eligibility & No (One Bitcoin addr. one vote) & Yes & Yes (Unclear how)
\tabularnewline
\hline
Privace & Yes & Yes & Yes 
\tabularnewline
\hline
Individual Verifiability & Yes & Yes & Yes 
\tabularnewline
\hline
Universal Verifiability & Yes & Yes & Yes 
\tabularnewline
\hline
Forgiveness & No & Yes (Unclear how) & Yes (Unclear how)
\tabularnewline
\hline

\end{tabularx}
\caption{Commercial blockchain e-voting protocols overview}
    \label{table:CommercialEvoteOverview}
\end{minipage}
\end{table}

\section{Proposed Protocol}
\label{sec:ProposedProtocol}
The motivation behind the proposed e-voting protocol,  is to have a blockchain based scheme that meets the above stated goals. In addition to those properties the protocol must allow for a voter to change one's mind and cancel one's vote, replacing it with another.

As a secondary goal, it has been actively pursued to provide the maximum degree of decentralisation and to create a protocol which the voters control as a network of peers. After careful consideration, however, it was decided that a certain degree of centralisation is necessary to reach the primary goal. This is because when using the blockchain, one is unable to store secret information in the public ledger without the use of external oracles that maintain such information. So if the identity of the voters is to remain secret, whilst at the same time permitting only eligible voters to participate in the elections, a Central Authority needs to be introduced that acts as a trusted third party.

The proposed voting protocol utilises the blockchain to store the cast ballots, therefore in this context the blockchain acts as a transparent ballot box. The main reason for using the blockchain in an e-voting protocol is to take advantage of the fact that it enables a group of people to maintain a public database, that is owned, updated, and maintained by every user, but controlled by no one. Since the protocol is based on the blockchain, it will be realised as a network of peers. Each voter will be a peer i.e. a node in a network of equals. Every voter will be responsible for making sure that fraudulent votes are rejected, hence that consensus is maintained according to the election rules. The blockchain also has the additional advantage of being increasingly well-known and well-trusted to operate as intended, as evidenced by the sheer size of the cryptocurrency market.

\subsection{Notations and Definitions}
During the description, of the protocol terms like voter, ballot and vote will be used. A list of all notations used, are also provided in Table \ref{tab:NotationTable}.

\begin{table}[h]
\caption{Notation used in protocol description.}
	\label{tab:NotationTable}
		\centering
\begin{tabular}{p{1cm} p{0.01cm} p{6.6cm}}\hline
$i$	&:&Voter $i$.\\
$V_{ipub}$	&:&Public key of voter $i$, uniquely identifying $V_i$, also serves as signature verifying key.\\
$V_{ipriv}$ &:& The private counterpart of $V_{ipub}$, used as a signing key.\\
$Sig_{CA}(X)$ &:& Digital signature produced by the CA over message x.\\
$C_i$ &:& Voter $i$'s choice in the election.\\
$dc_i$&:& Digital commitment for $C_i$.\\
$o_i$ &:& $dc_i$ opening value. $C_i$ cannot be derived from $dc_i$ without the opening value.\\
$x \vert y$ &:& Inclusion of value x and y in a single message\\
$et_i$ &:& Eligibility token, $et_i = Sign_{CA}(V_{ipub} \vert dc_i)$\\
$ $ &:& \\
\hline
\end{tabular} 
\end{table}

\subsubsection{Voter}
\label{sec:Voter}
A voter, identified by one's public key, $V_ipub$, is considered an entity that is permitted to cast a vote towards one of the candidates. A voter also has the ability to cast an Invalid vote as a form of protest, that will not be counted towards the end result, but will however, be logged in the system. Voters will be able to access the e-voting platform through a voting client, installed in their device of preference, the security of which is assumed. The terms client and voter will be used interchangeably to describe the entity acting as the voter.

The voter, during the election, will be called to make a choice, $C_i$ that can range from a set of predefined choices, to protest messages. In order to assure fairness, the voter will reveal only a digital commitment over said choice, $dc_i$ and will reveal the choice itself only during the counting stage of the election. 

\subsubsection{Central Authority (CA)}
\label{sec:CentralAuthority}
In order for the e-voting protocol to provide assurance that only eligible voters are able to vote, it was deemed necessary for a Central Authority to be introduced. For a user to be judged eligible, one must authenticate oneself to the Central Authority, CA, and receive a token that proves one's eligibility to vote. The eligibility token, eti, takes the form of a digital signature over a voters $V_{ipriv}$ and $dc_i$. 

\begin{equation}
 et_{i}= sig(V_{ipub} | dc_{i})_{CA}
\end{equation}

For the CA to judge whether voters are eligible or not, it is required for it to maintain a list of all the voters that are allowed to participate in the elections. . The CA is also required to have access to voter authenticating information in order to have the ability to authenticate eligible users. The method the CA will deploy to authenticate the voter, is application dependent. 

\subsubsection{Vote} 
A vote is a message of predefined structure, that is the equivalent of a bitcoin transaction. A vote is required to include a ballot and any other information that the practical implementation of the protocol requires it to include. Each vote x, once included in the blockchain is identified by a vote ID, $VID_x$, a value that uniquely recognises a vote. 

\subsubsection{Ballot} 
A ballot, bi is the digital representation of the physical ballot, i.e. the paper where the choice of a voter is written on. A ballot is considered sealed when the opening value of the digital commitment has not been revealed and, thus no party, other than the voter, can determine the way a voter voted. Once the opening value has been revealed and the choice of the voter is publicly known, the ballot is considered open.

\begin{equation}
b_ialt = V_{ipub}| dc_i | et_i
\end{equation}

A ballot is considered sealed when the opening value of the digital commitment has not been revealed and, thus no party, other than the voter, can determine the way a voter voted. Once the opening value, $o_i$, has been revealed and the choice of the voter is publicly known, the ballot is considered open. 

 The public key, included in the ballot denotes the owner of the ballot and by extension the owner of the vote. This ownership property enables the protocol to support a different kind of ballot namely the alteration ballot.

\subsubsection{Alteration Ballot} 
The alteration ballot serves the purpose of enabling voters to alter their vote, that has already been cast in the blockchain. The alteration ballot, $b_{ialt}$, is comprised of the following elements:

\begin{itemize}
\item A statement of the voter that has voted x, with unique ID, $VID_{x}$, will be cancelled, $c(VID_{x})$ and should not be counted towards the end result. Vote x, must already exist in the blockchain. 
\item The public key of the Voter, $V_{ipriv}$, that also acts as the signature validating key.
\item The new commitment of the voter, $dc_n$ encapsulating the voter's new choice. 
\end{itemize}

\begin{equation}
b_{i}= c(VID_{x}) | V_{ipriv} | dc_{n} | sig(c(VID_{x}) | V_{ipriv} | dc_{n})V_{i}
\end{equation}

\subsection{Voting Phases}
 The protocol has been divided in four distinct phases, namely the, initialisation phase, the preparation phase, the voting phase and finally the counting phase.
\subsubsection{Initialisation phase}
During the initialisation phase, the rules governing the elections are determined and the CA, the blockchain and all other systems of the protocol are initialised. The organisers of the elections will be called to decide, amongst others, on what the duration of the individual protocol phases will be and on whether vote cancelation will be permitted or not. The rules will then be publicised and a CA and a blockchain infrastructure will be created governed by those rules. 

The CA during the initialisation phase will be provided the list of the eligible voters that are eligible to vote as well as a way to authenticate those users. A pair of signing and verifying keys for the public signature scheme will be generated and the verifying key will be publicised as a system wide parameter. 

The blockchain will be initialised with an initialisation block, that will serve as the genesis block, of the chain. The initialisation block does not contain any votes, but instead it contains all the information of the election, including the CA's signature validating key, the set of valid choices the voters can choose from and so on. This way a blockchain is tied to a specific election and all the system parameters become part of the blockchain and thus dispute over them is prevented. A visual representation of the blockchain is depicted in Fig. 4s, where the election blockchain can be seen, comprised of singly linked blocks containing votes with unique VIDs, where the first block contains election specific information.

\begin{figure}[htpb]
 \centering
 \includegraphics[width=\columnwidth]{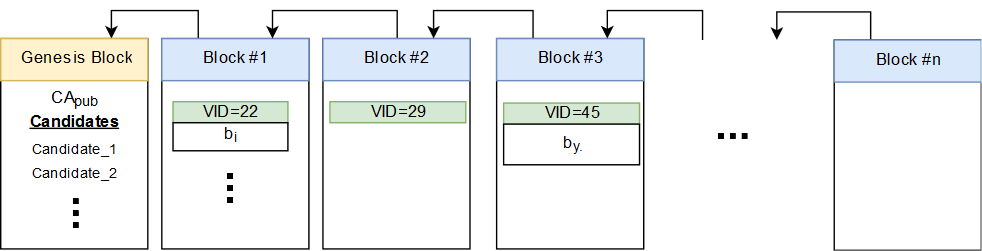}
 \caption{Vote alteration}
 \label{fig:Contracts} 
\end{figure} 

\subsubsection{Preparation phase}
During this phase, $V_i$, using the client application of the e-voting platform, is called to authenticate oneself to the Central Authority. The CA will use the list of eligible voters along with the authentication information, it acquired during the initialisation phase, to determine whether the aspiring voter is eligible to vote. If the voter is judged eligible, the CA will proceed to the following steps, otherwise $V_i$ is rejected and the CA, does not proceed with the rest of the phase. All the following information will be exchanged through an authenticated and secure chanel. financial transactions , over an unreliable channel. 

Once deemed eligible, the $V_{i}$'s client will generate a public key pair, whose public counterpart $V_{ipub}$, will be used as a pseudonymous identity of the voter and will also serve as one's verifying key. It will also prompt the voter to make one's choice, $c_{i}$ of the predetermined choices that will be accepted by the system and a digital commitment scheme will be used in order to generate $dc_{i}$, the digital commitment of one's choice.

To prove one's eligibility to the system the voter needs to send both $dc_{i}$ and $V_{ipub}$ to the CA to be signed by it. To avoid the possibility of the CA linking a voter's true identity to ones vote a blinding signature scheme will be used. The client will apply a blinding function on the message, blind($V_{ipub} | dc_{i}$) and send it to the CA that will then sign the blinded message and send it back. Once the message is received, the client will unblind the signed message $sig^{-1}(sig(V_{ipub}|dc_i)CA)= sig(V_{ipub}|dc_{i})CA$ and will end up with a valid eligibility token $et_i$.

\subsubsection{Voting phase}
During the voting phase, every Voter constructs and then broadcasts to the network their vote. Each voter is also responsible for collecting votes, validating them and inserting the valid ones in the blockchain. In order for a voter to accept a vote as a valid one and include it in a block, one will make sure that the owner of the vote has not previously cast that vote. One will also have to make sure that CA’s signature included in the ballot is validated and that the vote adheres to the predefined structure. If any of those checks fail the vote is discarded as an invalid one. 

The protocol provides the option to allow voters to change their vote even after those votes have been cast, through the use of the alteration ballot $b_{ialt}$. Since the alteration ballot includes a valid signature over the new ballot and the same public key validates both the alteration ballot and the previously cast vote, the network will accept bialt, include it in the blockchain and proceed with including it in the count instead of the cancelled one. Ballot cancelation can be performed multiple times and only the final ballot will be included in the count. 

\subsubsection{Counting phase} 
During the counting phase, all voters are called to reveal their final choice by broadcasting to the network a ballot opening message, $ob_{i}$, \ containing the VID of their final vote in the blockchain, the opening value of their vote commitment, and a signature over both values. 

\subsubsection{Counting phase} 
During the counting phase, all voters are called to reveal their final choice by broadcasting to the network a ballot opening message, obi, \ containing the VID of their final vote in the blockchain, the opening value of their vote commitment, and a signature over both values. 

\begin{equation}
ob_{i}=VID_{x} | o_{i} | sig(VID_{x} | o_{i})CA
\end{equation}

All nodes of the network will be responsible for collecting the ballot opening messages and verifying that the signature validates with the public key of the owner of vote $VID_x$. If the signature is verified, the voters will then broadcast the messages to their adjacent peers. And proceed with including the vote in their count. All peers should reach the same result since they operate on the same blockchain.

\subsection{Protocol Analysis}
In the following paragraphs, the extent to which the protocol satisfies the e-voting design properties will be examined, in order determine if the property is held.

\subsubsection{Eligibility} 
In order for a vote to be included in the blockchain and thus considered cast it needs to include a valid ballot. Each ballot needs to include a valid signature of the CA, over V\textsubscript{ipub} and dc\textsubscript{i}, otherwise it is dropped by the network. The CA provides signatures only to authenticated voters that have been included in the list of eligible voters compiled during the initialisation phase and that haven't requested to vote before. \ This means only eligible voters can vote and they can acquire only one eligibility token and thus cast only one valid vote. \ 

 The eligibility property also breaks when an eligible voter succeeds in casting a vote more than once. This, however, is not possible since all nodes, during the voting phase, will refuse to include to the blockchain a vote that has already been cast. The only vote accepted by the network after V\textsubscript{i} has voted will be an alteration vote and since it bears a valid signature of V\textsubscript{i} only the owner of the vote to be canceled can produce. The vote alteration process can be better illustrated in \ref{fig:Contracts} where a voter, V\textsubscript{i}, with public key $V_{ipriv}$, just cast an alteration vote with VID=45. That new vote cancels another, previously cast, alteration vote with VID=29 that in turn cancelled V\textsubscript{i}’s original vote, VID=22.

Since the original vote includes proof, issued by the CA, that the holder of public key $V_{ipriv}$ is eligible to vote and both vote 29 and 45 have, provably, been issued by the same entity, vote 45, that will be counted towards the final tally, is the vote of an eligible voter.

\subsubsection{Privacy} 
The protocol guarantees that at no point during the protocol run, any party can determine how a voter voted. \ The only link between the real identity of the voter and one's vote is an individual's public key that acts as a pseudonymous identity. The only entity that would potentially uncover said link would be the CA, since it is the only entity that the voter would need to reveal one's true identity to in order to obtain proof of one's eligibility, during the preparation stage. 

In order to avoid any party from being able to identify how a voter voted, a blind signature scheme is used. Blind signatures, provide a way for the central authority to produce a valid signature on the digital commitment and public key of a voter, without being able to determine neither the public key nor the digital commitment.

The privacy of the voter is maintained, even with the addition of the voting alteration mechanism since the fact remains that the only entity that can link a public key with a real-life identity is the CA and the CA does not gain any additional knowledge with the introduction of that mechanism.

\subsubsection{Fairness} 
The protocol provides a guarantee that the results will not be known during the voting phase and, thus, no voters will be swayed. This is achieved through the use of a digital commitment scheme and by separating the voting stage from the counting stage. The voters during the first three stages of the protocol will not make their choices known but will instead include in the ballot a digital commitment of said choice. The ballot will be opened, only during the counting phase, when the voters will reveal their choice by broadcasting the opening value of the digital commitment. If votes are revealed before the counting phases, they should not count towards the end result.

\subsubsection{Individual verifiability}
Due to the public nature of the ledger, each voter can verify that one's vote has been inserted in the blockchain, thus has been counted. Each voter is also responsible for counting the votes and thus one can ensure that the result includes one's vote.

\subsubsection{Universal verifiability}
Since the ledger is public, every voter can verify that the votes have been counted correctly, by simply counting the votes. External auditors can also verify the results by obtaining a copy of the blockchain, making sure that the votes in it are legitimate, e.g. that signatures are validated, duplicates don't exist etc. and once all checks are complete, auditors can count the votes and compare their results against the official election tally. The fact that rules governing the election are included in the genesis block of the blockchain, further facilitates the election's verifiability since their integrity is guaranteed and thus disputes over them become irrelevant.

\subsection{Further considerations}

The CA, is the only centralisation point of the protocol and it is assumed to be trusted. However, if the CA breaks that trust in the current setting, it could arbitrarily cast votes for voters that haven't voted. If the CA would not surpass the number of voters that participate in the election, those fraudulent votes could not be detected. This is why all voters should cast their vote. An additional failsafe would be to introduce a multisignature scheme where more than one independent CAs, would need to sign an eligibility token in order to produce a valid signature over it. \ Each CA would maintain only part of a voters' authenticating information, thus making it impossible for one to impersonate another voter. It should be noted that the introduction of additional CAs does come with an overhead and it is something that should be considered. 

Since only voters are allowed to participate in the elections, the suggested blockchain to be used is a Permissioned one. The voters can use their eligibility tokens, as proof that they can participate in the blockchain. This makes the environment inherently more secure.

\section{Implementation and Operational Evaluation}
\label{sec:ImplementationandOperationEvaluation}
In this section, we will give an example of an implementation of this proposed protocol.

\subsection{Implementation Details}
\label{sec:ImplementationDetails}
For this implementation, we have decided to use a private network that uses the Ethereum blockchain API. The reasoning behind this decision is that Ethereum is a widely recognised and popular technology. Comparative protocols are likely also implemented on Ethereum, thus giving a better basis of comparison with protocols with similar goals. 

When creating applications that use Ethereum, computational expenses impacts  the design choices. With Ethereum, this computational expense is manifested in the form of 'gas'. Gas is a unit of measure that decided the computational expense of a contract. Gas is priced by the node wishing to push the node to the wider chain, and this price will be paid to the node that mines that transaction. Nodes, therefore, are attempting to maximise profits by determining the worth of a transaction verses the computational cost. As a result, to make a blockchain application viable (and to ensure that their transactions aren't always being past over by the mining nodes), the computational expense has to be minimised. This is less true in a private network. By merit of being a part of a private network, we can assume that the nodes have other stakes in the chain. In this instance, that is the right to vote. Still, computation still has attached expenses. There are ethical complications in charging voters for the right to vote, particularly if that cost is high. A means to circumvent this would be to have a gateway node that is sponsored in some way by the election hosts. This node could then be used by voters without the means to join the private network by staking computational power. This needs to be balanced by a sufficient number of nodes participating. Nevertheless, the need to minimise computational expense has played a role in the design decisions made.

\begin{algorithm}
\caption{Initialisation Phase}\label{alg:initiateElection}
\addtolength\linewidth{-4ex}
\small
\begin{algorithmic}[1]
\Procedure{ElectionGenesis}{\emph{\_candidates},\emph{\_pubk},\emph{\_lengthPhaseOne},\emph{ \_lengthPhaseTwo},\emph{\_cancelBallots}}
\State $candidates\gets \_candidates$
\State $pubk\gets \_pubk$
\State $electionEndTime\gets timeNow() + \_lengthPhaseOne$
\State $countEndTime\gets electionEndTime + \_lengthPhaseTwo$
\State $cancelBallots\gets \_cancelBallots$
\EndProcedure
\end{algorithmic}
\end{algorithm}

\begin{table}[htb]
\centering
\begin{tabular}{||c c c||} 
 \hline
 Contract & Time (secs) & Cost (gas) \\ [1ex] 
 \hline\hline
 1 & 50 & 577207 \\ 
 \hline
 2 & 53 & 577207  \\
 \hline
 3 & 55 & 577207  \\
 \hline
 4 & 54 & 577207 \\
 \hline
 5 & 56 & 577207 \\  
 \hline
 6 & 55 & 577207 \\ 
 \hline
 7 & 59 & 577207  \\
 \hline
 8 & 56 & 577207 \\
 \hline
 9 & 56 & 577207 \\
 \hline
 10 & 54 & 577207 \\ 
 \hline\hline
 Average & 54.8 & 577207 \\ 
 \hline
\end{tabular}
\caption{Election Genesis: Contract Deployment}
\label{table:ElectionGenesisContractDeployment}
\end{table}

\subsubsection{Initialisation Phase}
During the initilisation of an election, a genesis contract must be placed on the blockchain (Algorithm \ref{alg:initiateElection}). This genesis contract contains all of the information that will be necessary to validate the ballots as they are placed, ensuring that none are placed after time, that none are placed without the appropriate signed token, and that alteration ballots are not cast if the election forbids such ballots. This contract will also contain other information that should be publicly available to everyone participating in the election. Such information includes the list of candidates that can be voted for and the timing for all of the phases of the election.

When the contract is created, this information will be passed in by the client responsible for pushing the contract to the blockchain. It will then be stored in the contract state.

\subsubsection{Voting Phase - Initial Ballot}
In order to place a ballot on the blockchain, (Algorithm \ref{alg:placeInitialBallot}) the voter must have first communicated with the CA to receive a signed token authorising the ballot. When they make their submission, the voter must include the component parts of this signed token (separation of the token into it's component parts costs too much gas on the blockchain to be considered feasible). 

\begin{algorithm}[htb]
\caption{Voting Phase - Initial Ballot }\label{alg:placeInitialBallot}
\addtolength\linewidth{-4ex}
\small
\begin{algorithmic}[1]
\Procedure{PlaceBallot}{\emph{vid},\emph{vote},\emph{msghashed},\emph{ v},\emph{r},\emph{s}}
\State \textbf{require}((\emph{timeNow()} $<$ \emph{electionEndTime}) \textbf{And} (\emph{verifyToken}(\emph{msghashed},\emph{v},\emph{r},\emph{s}))
\State \textbf{new} InitialBallot(vid, vote)
\EndProcedure
\Procedure{InitialBallot}{\emph{\_vid},\emph{\_vote}}
\State $vid\gets \_vid$
\State $vote\gets \_vote$
\State $sealed\gets true$
\State $unsealedTimeStamp\gets null$
\EndProcedure
\end{algorithmic}
\end{algorithm}

Along with the candidate that they have voted for, the voter must also submit a VID that will be used to uniquely mark this vote. This is to prevent a voter from placing multiple bids on the blockchain (where more than one is not an altering ballot).

\begin{figure}[ht]
	\centering
\centering
\begin{tikzpicture}
  \begin{axis}
    [ylabel = {Time (Seconds)},
    boxplot/draw direction=y,
    xtick={1,2,3, 4, 5, 6, 7, 8, 9, 10},
    xticklabels={T1, T2, T3, T4, T5, T6, T7, T8, T9, T10},
    x tick label style={font=\footnotesize, text width=2.5cm, align=center}
    ]
    \addplot+[
    boxplot prepared={
      lower whisker=71,
      lower quartile=74.25,
      median=78.5,
      upper quartile=79.75,
      upper whisker=81,
      average=76.9
    }, color = red
    ]coordinates{};
    \addplot+[
    boxplot prepared={
      lower whisker=58,
      lower quartile=68,
      median=80.5,
      upper quartile=80.75,
      upper whisker=90,
      average=76.4
    }, color = blue
    ] coordinates{};
    \addplot+[
    boxplot prepared={
      lower whisker=50,
      lower quartile=53.75,
      median=56.5,
      upper quartile=73,
      upper whisker=82,
      average=62
    }, color = Aquamarine
    ] coordinates{};
    \addplot+[
    boxplot prepared={
      lower whisker=47,
      lower quartile=49.5,
      median=51,
      upper quartile=56.25,
      upper whisker=60,
      average=52.5
    }, color = Orange
    ]coordinates{};
    \addplot+[
    boxplot prepared={
      lower whisker=42,
      lower quartile=44.5,
      median=48.5,
      upper quartile=54,
      upper whisker=57,
      average=49.2
    }, color = OrangeRed
    ] coordinates{};
    \addplot+[
    boxplot prepared={
      lower whisker=44,
      lower quartile=45.5,
      median=48.5,
      upper quartile=51,
      upper whisker=58,
      average=49.3,
      every box/.style={thin,solid,draw=Periwinkle,fill=white},
      every whisker/.style={Periwinkle, thin, solid},
      every median/.style={solid,Periwinkle,thin}
    }, color = Periwinkle
    ] coordinates{};    
    \addplot+[
    boxplot prepared={
      lower whisker=39,
      lower quartile=42.25,
      median=43,
      upper quartile=43,
      upper whisker=47,
      average=42.9,
      every box/.style={thin,solid,draw=Purple,fill=white},
      every whisker/.style={Purple, thin, solid},
      every median/.style={solid,Purple,thin}
    }, color = Purple
    ]coordinates{};
    \addplot+[
    boxplot prepared={
      lower whisker=39,
      lower quartile=41.5,
      median=43.5,
      upper quartile=46.5,
      upper whisker=48,
      average=43.7,
      every box/.style={thin,solid,draw=CarnationPink,fill=white},
      every whisker/.style={CarnationPink, thin, solid},
      every median/.style={solid,CarnationPink,thin}
    }, color = CarnationPink
    ] coordinates{};
    \addplot+[
    boxplot prepared={
      lower whisker=42,
      lower quartile=43.5,
      median=47,
      upper quartile=47.75,
      upper whisker=54,
      average=46.5,
      every box/.style={thin,solid,draw=Maroon,fill=white},
      every whisker/.style={Maroon, thin, solid},
      every median/.style={solid,Maroon,thin}
    }, color = Maroon
    ] coordinates{};
       \addplot+[
    boxplot prepared={
      lower whisker=43,
      lower quartile=42.25,
      median=46.5,
      upper quartile=47,
      upper whisker=49,
      average=46.2,
      every box/.style={thin,solid,draw=JungleGreen,fill=white},
      every whisker/.style={JungleGreen, thin, solid},
      every median/.style={solid,JungleGreen,thin}
    }, color = JungleGreen
    ] coordinates{};
    \end{axis}
\end{tikzpicture}
\caption{Contract timing for Initial Vote Over Ten Trial Ballots}
\label{fig:IVtime}
\end{figure}
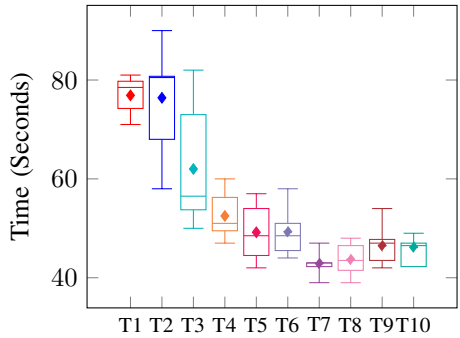

The message token is verified to have been signed by the public key belonging to this election, and the current time is checked to ensure that the vote is not being placed after the election end. 

Only after these requirements are met can the ballot be created. The VID and vote are set on this ballot contract. The ballot is also set to sealed and the unsealedtimeStamp, which is used to denote the time at which a vote was retrieved, is set to null as it has yet to be retrieved. This contract is then pushed to the blockchain.

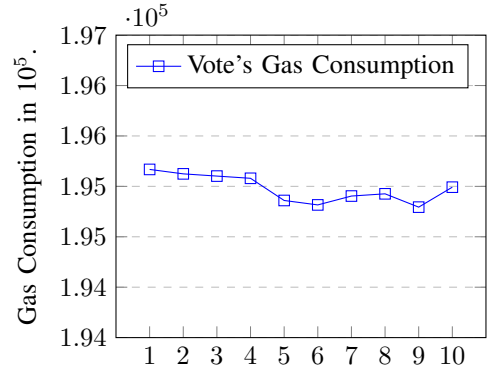
\begin{figure}[ht]
	\centering
\centering
\begin{tikzpicture}
\begin{axis}[
    ylabel={Gas Consumption in 10$^5$.},
    xmin=0, xmax=11,
    ymin=193500, ymax=196500,
    xtick={1,2,3,4,5,6,7,8,9,10},
    ytick={193500,194000,194500,195000,195500,196000,196500},
    legend pos=north west,
    ymajorgrids=true,
    grid style=dashed,
]
\addplot[
    color=blue,
    mark=square,
    ]
    coordinates {
    (1,195168)(2,195124)(3,195102)(4,195080)(5,194860)(6,194816)(7,194904)(8,194926)(9,194794)(10,194992)
    };
    \legend{Vote's Gas Consumption}
\end{axis}
\end{tikzpicture}
\caption{Individual Initial Vote's GAS usage (x 10$^5$)}
\label{fig:IVContractGasUsage}
\end{figure}

\subsubsection{Voting Phase - Alternate Ballot}
The process for pushing an altering ballot, (Algorithm \ref{alg:placeAlterBallot}) is similar to the process for creating an initial ballot. The voter must have a signed token authorising the ballot. When they make their submission, the voter must include the component parts of this signed token. Along with the candidate that they have voted for, and the unique VID, the candidate must also indicate the previous ballot that is being altered. The VID of this previous ballot must also include the address, to be extracted by the evaluating client code. The validation as to whether this is a legitimate altering ballot is not done until the counting phase. This decision was made, in part, because of the lack of support that Solidity (the language contracts on the Ethereum blockchain) has for extrapolating transactions from the blockchain. Moreover, the decision was made not to keep a store of the addresses of the previously placed ballots - it was too computationally expensive as this would involve altering state.

\begin{algorithm}
\caption{Voting Phase - Altering Ballot }\label{alg:placeAlterBallot}
\addtolength\linewidth{-4ex}
\small
\begin{algorithmic}[1]
\Procedure{PlaceAlterBallot}{\emph{vid},\emph{vote},\emph{msghashed},\emph{ v},\emph{r},\emph{s}}
\State \textbf{require}((\emph{timeNow()} $<$ \emph{electionEndTime}) \textbf{And} (\emph{verifyToken}(\emph{msghashed},\emph{v},\emph{r},\emph{s}) \textbf{And} cancelBallots)
\State \textbf{new} InitialBallot(vid, vote)
\EndProcedure
\Procedure{AlteringBallot}{\emph{\_vid},\emph{\_vote}, \emph{\_replacedBallot}}
\State $vid\gets \_vid$
\State $vote\gets \_vote$
\State $replacedBallot\gets \_replacedBallot$
\State $sealed\gets true$
\State $unsealedTimeStamp\gets null$
\EndProcedure
\end{algorithmic}
\end{algorithm}


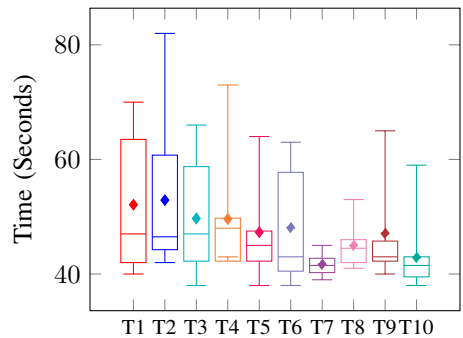
\begin{figure}[ht]
	\centering
\centering
\begin{tikzpicture}
  \begin{axis}
    [ylabel = {Time (Seconds)},
    boxplot/draw direction=y,
    xtick={1,2,3, 4, 5, 6, 7, 8, 9, 10},
    xticklabels={T1, T2, T3, T4, T5, T6, T7, T8, T9, T10},
    x tick label style={font=\footnotesize, text width=2.5cm, align=center}
    ]
    \addplot+[
    boxplot prepared={
      lower whisker=40,
      lower quartile=42,
      median=47,
      upper quartile=63.5,
      upper whisker=70,
      average=52.1
    }, color = red
    ]coordinates{};
    \addplot+[
    boxplot prepared={
      lower whisker=42,
      lower quartile=44.25,
      median=46.5,
      upper quartile=60.75,
      upper whisker=82,
      average=52.9
    }, color = blue
    ] coordinates{};
    \addplot+[
    boxplot prepared={
      lower whisker=38,
      lower quartile=42.25,
      median=47,
      upper quartile=58.75,
      upper whisker=66,
      average=49.7
    }, color = Aquamarine
    ] coordinates{};
    \addplot+[
    boxplot prepared={
      lower whisker=43,
      lower quartile=42.25,
      median=48,
      upper quartile=49.75,
      upper whisker=73,
      average=49.6
    }, color = Orange
    ]coordinates{};
    \addplot+[
    boxplot prepared={
      lower whisker=38,
      lower quartile=42.25,
      median=45,
      upper quartile=47.5,
      upper whisker=64,
      average=47.3
    }, color = OrangeRed
    ] coordinates{};
    \addplot+[
    boxplot prepared={
      lower whisker=38,
      lower quartile=40.5,
      median=43,
      upper quartile=57.75,
      upper whisker=63,
      average=48.1,
      every box/.style={thin,solid,draw=Periwinkle,fill=white},
      every whisker/.style={Periwinkle, thin, solid},
      every median/.style={solid,Periwinkle,thin}
    }, color = Periwinkle
    ] coordinates{};    
    \addplot+[
    boxplot prepared={
      lower whisker=39,
      lower quartile=40.25,
      median=41.5,
      upper quartile=42.75,
      upper whisker=45,
      average=41.7,
      every box/.style={thin,solid,draw=Purple,fill=white},
      every whisker/.style={Purple, thin, solid},
      every median/.style={solid,Purple,thin}
    }, color = Purple
    ]coordinates{};
    \addplot+[
    boxplot prepared={
      lower whisker=41,
      lower quartile=42,
      median=44.5,
      upper quartile=46,
      upper whisker=53,
      average=45,
      every box/.style={thin,solid,draw=CarnationPink,fill=white},
      every whisker/.style={CarnationPink, thin, solid},
      every median/.style={solid,CarnationPink,thin}
    }, color = CarnationPink
    ] coordinates{};
    \addplot+[
    boxplot prepared={
      lower whisker=40,
      lower quartile=42.25,
      median=43,
      upper quartile=45.75,
      upper whisker=65,
      average=47.1,
      every box/.style={thin,solid,draw=Maroon,fill=white},
      every whisker/.style={Maroon, thin, solid},
      every median/.style={solid,Maroon,thin}
    }, color = Maroon
    ] coordinates{};
       \addplot+[
    boxplot prepared={
      lower whisker=38,
      lower quartile=39.5,
      median=41.5,
      upper quartile=43,
      upper whisker=59,
      average=42.9,
      every box/.style={thin,solid,draw=JungleGreen,fill=white},
      every whisker/.style={JungleGreen, thin, solid},
      every median/.style={solid,JungleGreen,thin}
    }, color = JungleGreen
    ] coordinates{};
    \end{axis}
\end{tikzpicture}

\caption{Contract timing for Altering Vote Over Ten Trial Ballots}
\label{fig:AVtime}
\end{figure}

The message token is verified to have been signed by the public key belonging to this election, the current time is checked to ensure that the vote is not being placed after the election end, and, in this instance, the rules of the election are checked to ensure that altering ballots are permitted.

Only after these requirements are met can the ballot be created. The VID and vote are set on this ballot contract. The ballot is also set to sealed and the unsealedtimeStamp, which is used to denote the time at which a vote was retrieved, is set to null as it has yet to be retrieved. This contract is then pushed to the blockchain.

\begin{figure}[ht]
	\centering
\centering
\begin{tikzpicture}
\begin{axis}[
    ylabel={Gas Consumption in 10$^5$.},
    xmin=0, xmax=11,
    ymin=224000, ymax=227500,
    xtick={1,2,3,4,5,6,7,8,9,10},
    ytick={224500,225000,225500,226000,226500,227000,227500},
    legend pos=north west,
    ymajorgrids=true,
    grid style=dashed,
]
\addplot[
    color=blue,
    mark=square,
    ]
    coordinates {
    (1,225718)(2,225696)(3,225828)(4,225894)(5,225740)(6,225520)(7,225630)(8,225454)(9,225432)(10,225652)
    };
    \legend{Vote's Gas Consumption}
\end{axis}
\end{tikzpicture}
\caption{Individual Altering Vote's GAS usage (x 10$^5$)}
\label{fig:AVContractGasUsage}
\end{figure}
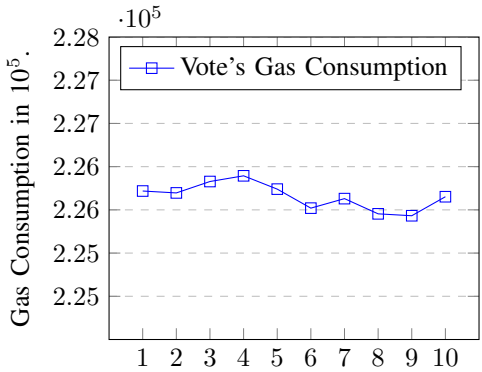\hspace{2mm}

\subsubsection{Counting Phase}
Once the election has concluded, votes will need to be counted (Algorithm \ref{alg:retrieveVote}). 

When the vote is retrieved from the ballot, if this is the first time that the vote has been retrieved, the ballot will be marked as unsealed. The time at which this retrieval was made is also recorded.

\begin{algorithm}
\caption{Counting Phase }\label{alg:retrieveVote}
\addtolength\linewidth{-4ex}
\small
\begin{algorithmic}[1]
\Procedure{RetrieveVote}{}
\State \textbf{require}(electionEndTime $<$ timeNow())
\If{$isSealed$}
\State $isSealed\gets false$
\State $unsealedTimeStamp\gets timeNow()$
\EndIf
\State \textbf{return}($vote$)
\EndProcedure
\end{algorithmic}
\end{algorithm}

The VID of the vote is publicly available. If this is an altering ballot, the replaced bid VID is also available. It is at this point that the altering bids can be checked for their validity before the vote is counted. 

Security is ensured by the fact that every node on the private blockchain has access to this information also and so can independently verify the final count.

\subsubsection{Challenging The Count}
Nodes on the blockchain also have functionality to examine the blockchain (Algorithm \ref{alg:retrieveSealed}). On top of the publicly readable information, such as the VID, the nodes can also explicitly and easily retrieve information such as the current sealed state of a ballot and the time at which the ballot was unsealed. If they have reason to contest whether a ballot was opened early, or not included in the count, the information given here may support the argument.

\begin{algorithm}
\caption{Challenging Count}\label{alg:retrieveSealed}
\addtolength\linewidth{-4ex}
\small
\begin{algorithmic}[1]
\Procedure{ReturnSealed}{}
\State \textbf{return}($isSealed$)
\EndProcedure
\Procedure{ReturnTimeUnsealed}{}
\State \textbf{return}($unsealedTimeStamp$)
\EndProcedure
\end{algorithmic}
\end{algorithm}


\section{conclusion}
\label{sec:conclusion}
E-voting, as discussed in the paper, is a potential solution to the lack of interest in voting amongst the young tech savvy population. For e-voting to become more open, transparent, and independently auditable, a potential solution would be base it on blockchain technology. This paper explores the potential of the blockchain technology and its usefulness in the e-voting scheme. The paper proposes an e-voting scheme, which is then implemented. The implementation and related performance measurements are given in the paper along with the challenges presented by the blockchain platform to develop a complex application like e-voting. The paper highlights some shortcomings and presents two potential paths forward to improve the underlying platform (blockchain technology) to support e-voting and other similar applications. Blockchain technology has a lot of promise; however, in its current state it might not reach its full potential. There needs to be concerted effort in the core blockchain technology research to improve is features and support for complex applications that can execute within the blockchain network.




\bibliographystyle{IEEEtran}
\bibliography{blockchain_paper.bib} 

\end{document}